\documentclass[aps,prd,preprint,groupedaddress,showkeys,showpacs,amsmath]
              {revtex4}

\newcommand{\ele}{{\mathrm e}}
\newcommand{\neu}{{\mathrm n}}
\newcommand{\pro}{{\mathrm p}}

\newcommand{\li}{{\mathrm L}{\mathrm i}}

\newcommand{\stat}{{\mathrm s}{\mathrm t}{\mathrm a}{\mathrm t}}
\newcommand{\sys}{{\mathrm s}{\mathrm y}{\mathrm s}}
\newcommand{\stor}{{\mathrm s}{\mathrm t}{\mathrm o}{\mathrm r}{\mathrm a}
                   {\mathrm g}{\mathrm e}}
\newcommand{\loss}{{\mathrm l}{\mathrm o}{\mathrm s}{\mathrm s}}

\usepackage{graphicx}
\usepackage{dcolumn}
\usepackage{bm}

\begin{document}

\title{Inhomogeneous Big Bang Nucleosynthesis Revisited}

\author{Juan F. Lara}
\email{ljuan@clemson.edu}
\affiliation{Department of Physics and Astronomy, 
             Clemson University,
             Clemson, South Carolina 29631, USA}
\affiliation{National Astronomical Observatory of Japan,
             2-21-1 Osawa, Mitaka,  
             Tokyo 181-8588, Japan}
\author{Toshitaka Kajino}
\email{kajino@nao.ac.jp}
\affiliation{National Astronomical Observatory and Graduate University for 
                Advanced Studies, 
             2-21-1 Osawa, Mitaka,  
             Tokyo 181-8588, Japan}
\affiliation{Department of Astronomy, Graduate School of Science, 
             University of Tokyo,
             7-3-1 Hongo, Bunkyo-ku
             Tokyo 113-0033, Japan}
\author{Grant J. Mathews}
\email{gmathews@nd.edu}
\affiliation{University of Notre Dame,
             Center for Astrophysics,
             Notre Dame, IN 46556, USA}
\affiliation{National Astronomical Observatory of Japan,
             2-21-1 Osawa, Mitaka,  
             Tokyo 181-8588, Japan}
\date{\today}

\begin{abstract}
We reanalyze the allowed parameters for inhomogeneous big bang nucleosynthesis
in light of the WMAP constraints on the baryon-to-photon ratio $\eta$ and a 
recent measurement which has set the neutron lifetime to be 878.5 $\pm$ 0.7 
$\pm$ 0.3 seconds.  For a set baryon-to-photon ratio $\eta$ the new lifetime 
reduces the mass fraction of $ ^{4}$He by 0.0015 but does not significantly 
change the abundances of other isotopes.  This enlarges the region of 
concordance between $ ^{4}$He and deuterium in the parameter space of $\eta$ 
and the IBBN distance scale $r_{i}$.  The $ ^{7}$Li abundance can be brought 
into concordance with observed $ ^{4}$He and deuterium abundances by using
depletion factors as high as 9.3.  The WMAP constraints, however, severely 
limit the allowed comoving ($T = $ 100 GK) inhomogeneity distance scale to
$r_{i} \approx (1.3-2.6) \times 10^{5}$ cm.
\end{abstract}

\pacs{26.35.+c}

\keywords{Inhomogenous big bang nucleosynthesis, Neutron Lifetime, Cosmic 
          Microwave Background}

\maketitle

\section{Introduction}

Big bang nucleosynthesis (BBN) plays a crucial role in constraining our views 
on the universe.  It is essentially the only probe of physics in the early 
radiation dominated epoch during the interval from $\sim 1-10^4$ sec.
Moreover, big bang cosmology is currently undergoing rapid evolution based upon
high precision determinations of cosmological parameters and improved input 
physics.  Thus, it is important to scrutinize all possible variants of BBN and 
understand the constraints which modern observations place upon their 
parameters.  This paper summarizes the current status on one such important
variant of the big bang, namely one in which the baryons are spatially 
inhomogeneously distributed during the epoch of nucleosynthesis.  We consider 
the constraints  from {\it WMAP} and the latest observed elemental abundances 
and include recent measurements of thermonuclear reaction rates and the neutron
lifetime.  We show, that although the parameter space for inhomogeneous big
bang nucleosynthesis (IBBN) is significantly limited,  interesting regions 
remain and this paradigm continues to be a viable possibility for the early 
universe.

\section{Background}

At a high temperature in the early universe, e.g. $T \sim $ 100 GK (the 
corresponding age of the universe is $\sim$ 0.01 seconds) baryons mainly exist 
in the form of free neutrons and protons.  During this epoch neutrons and 
protons are rapidly interconverted via the following weak reactions 
\cite{wagoner73,ytsso84,skm93}

\begin{eqnarray*}
   \neu + \nu_{\ele} & \leftrightarrow & \pro + \ele^{-} \\
     \neu + \ele^{+} & \leftrightarrow & \pro + \bar{\nu}_{\ele} \\
                \neu & \leftrightarrow & 
                            \pro + \ele^{-} + \bar{\nu}_{\ele}.
\end{eqnarray*}

As long as these reactions are in thermal equilibrium the ratio of neutrons to 
protons is given by $( \neu/\pro ) = \exp [ -\Delta m / kT ]$ where $\Delta m$
is the mass difference between neutrons and protons.  When the universe cools 
to a temperature $T = $ 13 GK these weak reactions fall out of equilibrium.  
The n/p ratio after that time decreases only due to neutron decay.

When the universe cools to a temperature $T \approx $ 0.9 GK the nuclear 
reaction n + p $\leftrightarrow$ d + $\gamma$ falls out of nuclear statistical
equilibrium.  Deuterium production becomes significant, leading to the 
synthesis of increasingly heavier nuclei through a network of nuclear 
reactions and weak decays.  When nearly all free neutrons are incorporated into
$ ^{4}$He nuclei, $ ^{4}$He production virtually stops.  The final mass 
fraction can be approximated as

\begin{eqnarray}
   Y_{p} & \approx & 2( \neu / \pro ) / [ ( \neu / \pro ) + 1 ].
\end{eqnarray}

Detailed final abundances depend most sensitively on the baryon-to-photon ratio
$\eta$.  Observational measurements of $ ^{4}$He, deuterium and $ ^{7}$Li 
abundances can be compared with abundances calculated from BBN to put 
constraints on $\eta$.  Constraints can also be put on $\eta$ from measurements
of the Cosmic Microwave Background (CMB).  The physics of accoustic 
oscillations in the CMB angular power spectrum has a dependence on various 
cosmological parameters (See Melchiorri \cite{melchiorri03} for comprehensive 
references to methods of CMB calculations), including the baryon density factor
$\Omega_{b} h^{2}$, and therefore $\eta$.  Recent CMB measurements using 
{\it WMAP} \cite{bennett03} set $\Omega_{b} h^{2} = 0.0224 \pm 0.0009$ and 
$\eta = 6.13 \pm 0.25 \times 10^{-10}$.  See Lara \cite{lara05} for references 
to previous attempts to constrain $\eta$ using BBN or the CMB.

The final n/p ratio depends on the neutron lifetime, $\tau_{\neu}$, which 
partially determines when weak reactions fall out of equilibrium.  The most 
recent lifetime world average \cite{eidelman04} is 
$\tau_{\neu} = 885.7 \pm 0.8$ seconds.  But a recent measurement by Serebrov et
al. \cite{serebrov05} sets $\tau_{\neu} = 878.5.\pm 0.7_{\stat} \pm 0.3_{\sys}$
seconds, which differs from the world average by six standard deviations.  
Serebrov et al. measured the storage loss rate 
$\tau^{-1}_{\stor} = \tau^{-1}_{\neu} + \tau^{-1}_{\loss}$ of ultracold 
neutrons in a gravitational trap, and then extrapolated the data down to 
$\tau^{-1}_{\loss} \rightarrow 0$ to get a value for $\tau^{-1}_{\neu}$ alone.
Their storage loss rate was as much as a factor of two smaller than 
measurements from previous experiments.  So their measurements were much closer
to the true neutron decay rate, greatly reducing the sytematic error in the 
measurements.  Also, Serebrov et al. used the neutron lifetime as a unitarity 
test of the Cabibbo Kobayashi Maskawa (CKM) matrix.  The Standard Model of 
particle physics predicts that the CKM matrix is unitary 
\cite{aabdgmnrz02,abdghmss04}.  The neutron lifetime by Serebrov et al. was in
better agreement with this unitarity prediction than the world average 
lifetime, as shown in Figure 4 of Serebrov et al. \cite{serebrov05}.

Mathews et al. \cite{mks05} applied the measurement $Y_{p} = 0.2452 \pm 0.0015$
from Izotov et al. \cite{icfggt99}, which is derived from observations of 
extragalactic HII regions in low metallicity irregular galaxies, to a Standard 
BBN (SBBN) code \cite{wagoner73,kawano92} available for public download 
\cite{website}.  In the results using the world average neutron lifetime, the 
above $1\sigma$ range of $Y_{p}$ corresponds to a range 
$\eta = 4.8 \pm 0.8 \times 10^{-10}$, not in agreement with the {\it WMAP} 
results.  Mathews et al. \cite{mks05} put in the Serebrov et al.lifetime value 
into the SBBN code \cite{website}.  The new lifetime has the effect of lowering
$Y_{p}$ by $\approx$ 0.0015 for all values of baryon-to-photon ratio $\eta$.
The $1\sigma$ range of $\eta$ corresponding to $Y_{p} = 0.2452 \pm 0.0015$ 
\cite{icfggt99} shifts to $\eta = 5.5 \pm 0.9 \times 10^{-10}$.  The new range 
of $\eta$ is in agreement with the range $\eta = 6.13 \pm 0.25 \times 10^{-10}$
 from the {\it WMAP} measurements.

This article will discuss how the measured isotope abundances and the 
{\it WMAP} data constrain parameters from an IBBN cosmology when using the 
Serebrov et al. neutron lifetime.  Various theories of cosmic phase transitions
may lead to the formation of baryon inhomogeneous regions, for example during a
first order quark hadron phase transition 
\cite{witten84,kurkisuonio88,mm93,skam90,afm87,ts89,kll01,bg92,ct93} or during 
the electroweak phase transition \cite{fjmo94,heckler95,bdr95,kks99,ma05}.  
Inhomogeneities can have planar, cylindrical or spherical symmetry 
\cite{okbm97,kks99,lss01,lss03,ma05}.  The evolution of baryon inhomogeneous 
regions can be modelled using an IBBN code 
\cite{zeldovich75,ep75,hogan78,bm83,sm86,ahs87,afm87,mf88,kmcrw88,ts89,km89,km90,kmos90,kmf90,mmaf90,mamf91,tsommf94,jfmk94,jfm94,jmf95,ls96,mko96,kjm97,sm98,is01,jr01,ks01,kajino02,keihanen02,mdns04,mfnhs05,lara05}
The inhomogeneous region is broken up into 64 zones, using a stretching 
function by Kurki-Suonio and Matzner \cite{km90}.  Within each zone $s$ the 
time evolution of the number density $n(i,s)$ of isotope species $i$ obeys the 
following equation \cite{km90,mmaf90,kks99,lara05}

\begin{eqnarray}
   \frac{\partial n(i,s)}{\partial t} & = & n_{b}(s) \sum_{j,k,l} N_{i} \left
         ( -\frac{Y^{N_{i}}(i,s) Y^{N_{j}}(j,s)}{N_{i}!N_{j}!}[ij] + 
          \frac{Y^{N_{k}}(k,s) Y^{N_{l}}(l,s)}{N_{k}!N_{l}!}[kl] \right ) 
         \nonumber  \\
   &   & -3\frac{\dot{a}}{a}n(i,s) + \frac{1}{r^{p}}\frac{\partial}{\partial r}
           \left ( r^{p} D_{n} \frac{\partial \xi}{\partial r}
                   \frac{\partial n(i,s)}{\partial \xi} \right ).
   \label{eq:dndt}
\end{eqnarray}

\noindent The first two terms correspond to nuclear reactions and weak decays 
that create or destroy isotope $i$ within zone $s$.  The weak reactions that 
interconvert neutrons and protons go in these terms for instance.  The third 
term coresponds to the expansion of the universe.  The final term corresponds 
to diffusion of isotope $i$ between zones.  Neutron diffusion is the most 
prominent diffusion because of the neutrons' lack of charge.  The influence of 
neutron diffusion on IBBN results is described in detail by Lara \cite{lara05}.
In both Lara \cite{lara05} and this article the authors use expressions derived
by Jedamzik and Rehm \cite{bc91,kagmbcs92,jr01,jedamzik04} for the diffusion 
coefficients.  Thermonuclear reaction rates from Angulo et al. \cite{angulo99} 
were used in runs for both Lara \cite{lara05} and this article.

\section{Results}

The following results are for a cylindrically symmetric IBBN model with an 
initial high baryon density in the outermost zones of the model.  Parameters
used to define the model include $R_{\rho}$, the initial ratio of the
high to low baryon densities, and $f_{v}$, the volume fraction of the high 
density region.  $R_{\rho}$ is set to $10^{6}$, and $f_{v}$ is defined such 
that the radius of the low density cylindrical core equals 0.925 of the total
radius (the high density outer shell then has a thickness equal to 0.075).  
This model was used by Orito et al. \cite{okbm97} and by Lara 
\cite{lara04,lara05} to optimally satisfy the light element abundance 
constraints and also the inhomogeneous geometry which is calculated in an 
effective kinetic model of QCD \cite{skam90}.  Final abundance results are 
presented as contour maps which are dependent on $\eta$ and the inhomogeneity 
distance scale $r_{i}$, the size of the outermost zone at the initial 
temperature of 100 GK.  For the smallest values of $r_{i}$, neutron diffusion 
homogenizes both neutrons and protons early enough for the resulting abundances
to have the same values as abundances from SBBN models.  

Note that the code used by Mathews et al. \cite{website} uses Bessel function 
expansions \cite{wagoner73} to solve for the rates of the neutron proton 
interconversion reactions \cite{weinberg72}.  The code used in this article 
uses a numerical integration method to solve for those rates \cite{lara01}.  
The difference in the methods of calculating the rates can lead to 
discrepencies in the value of $Y_{p}$ on the order of 0.001.  There are also 
differences in the calculation of the Coulomb correction and the zero 
temperature radiative correction between the codes that affect the results
\cite{dkgstt82,lara01}.  These differences will be discussed and resolved in a 
subsequent article.

\begin{figure}
   \includegraphics[scale=0.80,angle=90]{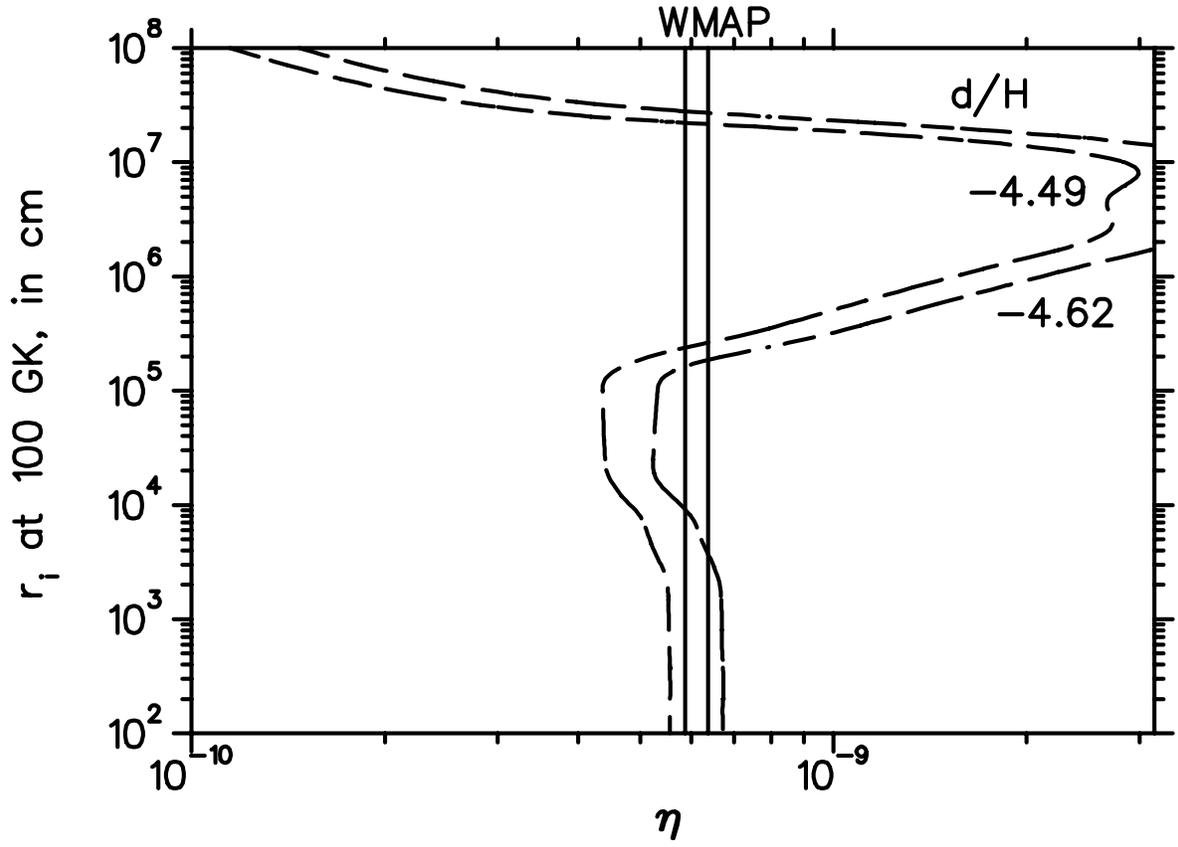}   
   \caption{\label{fig:deucmb} Contours consistent with the $\pm 2\sigma$ 
            measured abundance ratio D/H of deuterium \cite{ktsol03} are shown 
            in long dashed lines.  The CMB constraints from the {\it WMAP} 
            measurements \cite{bennett03} are also shown as solid lines.  The 
            {\it WMAP} constraint limits the allowed parameter space to three 
            regions.}
\end{figure}

Figure~\ref{fig:deucmb} shows both observational constraints for deuterium in 
an $\eta$ vs. $r_{i}$ contour map and the CMB constraints 
$\eta = 6.13 \pm 0.25$ from the {\it WMAP} results \cite{bennett03}.  The 
deuterium observational constraints D/H = $2.78_{-0.38}^{+0.44} \times 10^{-5}$
are the average of measurements of six absorption-line systems towards five 
Quasi Stellar Objects \cite{ktsol03}.  In the SBBN model the deuterium 
constraints correspond to $\eta = (5.6-6.7) \times 10^{-10}$, encompassing the 
{\it WMAP} constraints.  As can be seen in Figure~\ref{fig:deucmb} the results 
for the smallest values of $r_{i}$ are equivalent to SBBN results.  The bends 
in the deuterium contour lines are explained in detail in Lara \cite{lara05}.
In models with $r_{i}$ from $\approx$ 25000 cm to $10^{5}$ cm, neutron 
diffusion occurs between weak freeze-out and nucleosynthesis.  Contour lines 
are shifted to lower $\eta$ (lower deuterium production) because 
nucleosynthesis is concentrated in regions with large proton density.  For 
$r_{i}$ from $\approx 10^{5}$ cm to $2.0 \times 10^{6}$ cm neutron back 
diffusion does not reach all regions of the model during the time of 
nucleosynthesis, and countour lines shift instead to higher $\eta$ (greater 
deuterium production).  At $r_{i} \approx 2.0 \times 10^{6}$ cm neutron 
diffusion peaks at the same time as nucleosynthesis.  The deuterium contour 
lines shift to lower $\eta$ and the IBBN model is approximately the average of 
a high baryon density SBBN model and a low density SBBN model.  Due to these 
bends the deuterium constraints are in concordance with the {\it WMAP} results 
for three ranges of distance scale:  $r_{i} \le $ 9000 cm which includes the 
SBBN case, $r_{i} = (1.3-2.6) \times 10^{5}$ cm and 
$r_{i} = (2.1-2.9) \times 10^{7}$ cm.

\begin{figure}
   \includegraphics[scale=0.80,angle=90]{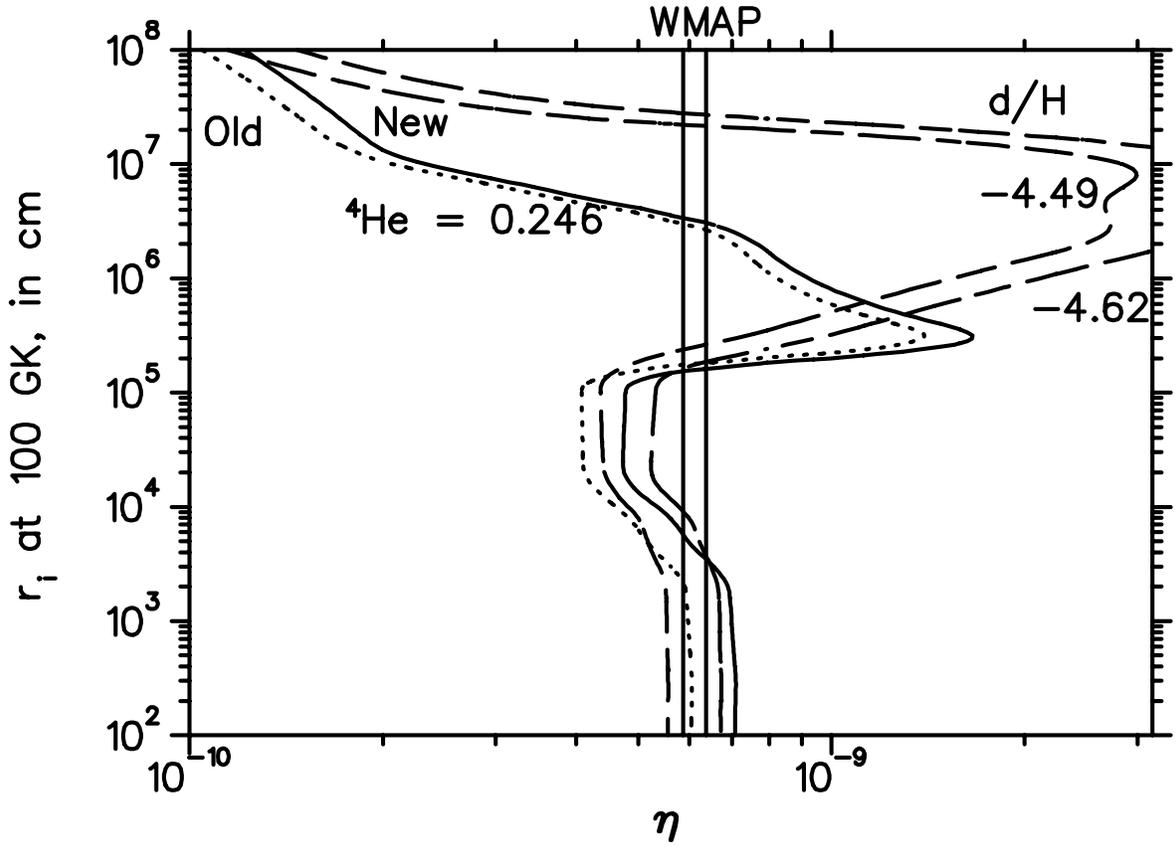}   
   \caption{\label{fig:deuhe4} Constraints from deuterium and CMB observations 
            are shown with the $+2\sigma$ maximum constraint of the $ ^{4}$He 
            mass fraction  $Y_{p}$ by Izotov and Thuan \cite{it04}.  
            The contour for the $+2\sigma$ maximum constraint of $Y_{p}$ is 
            shown by dashed line for the world average neutron lifetime 
            \cite{eidelman04} and solid line for the Serebrov et al. lifetime 
            \cite{serebrov05}.  The region to the right of the maximum 
            constraint is excluded.  The $+2\sigma$ minimum constraint 
            corresponds to values of $\eta$ so low it is not necessary to show 
            in this figure.  The lifetime effect is to shift the $Y_{p}$ 
            contour line to a higher $\eta$ by $\approx$ 1 $\times 10^{-10}$.}
\end{figure}

Figure~\ref{fig:deuhe4} shows observational constraints for $ ^{4}$He for the
cases of both the world average neutron lifetime and the neutron lifetime by
Serebrov et al., along with the deuterium and {\it WMAP} constraints.  
Figure~\ref{fig:deuhe4} shows $Y_{p} = $ 0.246, the $2\sigma$ maximum 
constraint of the measurement $0.242 \pm 0.002$ by Izotov and Thuan 
\cite{it04}.  This measurement is more recent than the previously adopted 
measurement by Izotov et al. \cite{icfggt99}.  If the world average neutron 
lifetime is used $Y_{p} = $ 0.246 corresponds to $\eta = 6.1 \times 10^{-10}$ 
for the cases of SBBN and IBBN with the smallest values of $r_{i}$.  Only a 
very narrow range $\eta = (5.88-6.1) \times 10^{-10}$ is then permitted for 
$r_{i} \le $ 2000 cm.  The region of concordance for 
$r_{i} = (1.3-2.6) \times 10^{5}$ cm however is contained within this 
$ ^{4}$He constraint.  Note though that the extent of systematic errors in 
$ ^{4}$He observations are controversial.  Olive and Skillman \cite{os04} adopt
a conservative approach to accessing uncertainties and report a very large 
range 0.232 $\le Y_{p} \le$ 0.258 due to correlated errors.  This range should 
eventually come down as the number of systems with uncorrelated systematic 
errors increase.

A shorter neutron lifetime means that more neutrons are converted to protons 
before the onset of nucleosynthesis, and $Y_{p}$ for a set value of $\eta$ is 
smaller.  But the lower neutron abundance has no perceptible effect ($< 1$ \%) 
in the final abundances of all other isotopes \cite{mks05}.  Since the neutron 
lifetime is independent of baryon inhomogeneity, the IBBN run shows the same 
results.  For a set $\eta$, $Y_{p}$ is reduced by 0.0015 while D/H is 
unchanged, regardless of the value of $\eta$ or $r_{i}$.  The contour line of 
the $ ^{4}$He maximum constraint \cite{it04} shifts to higher $\eta$ by 1 
$\times 10^{-10}$.  Using the world average neutron lifetime, the region of 
concordance between the deuterium and $ ^{4}$He constraints excludes a range
$(4.5 \times 10^{3}-1.3 \times 10^{5})$ cm of $r_{i}$.  The shift in the 
$ ^{4}$He contour line due to the Serebrov et al. lifetime allows for a
concordance between deuterium and $ ^{4}$He for all values of 
$r_{i} < 6.4 \times 10^{5}$ cm.  The {\it WMAP} constraints are in concordance 
with the deuterium constraints and $ ^{4}$He constraints with the Serebrov et 
al. lifetime for $r_{i} < $ 5500 cm.  The region of concordance for 
$r_{i} = (1.3-2.6) \times 10^{5}$ cm is still contained within the new 
$ ^{4}$He constraint.  For concordance with the region with 
$r_{i} = (2.1-2.9) \times 10^{7}$ cm a maximum value $Y_{p} = $ 0.259 is
needed.  This value is close to the maximum limit reported by Olive and 
Skillman, but that maximum limit is expected to diminish as the errors improve.

\begin{figure}
   \includegraphics[scale=0.80,angle=90]{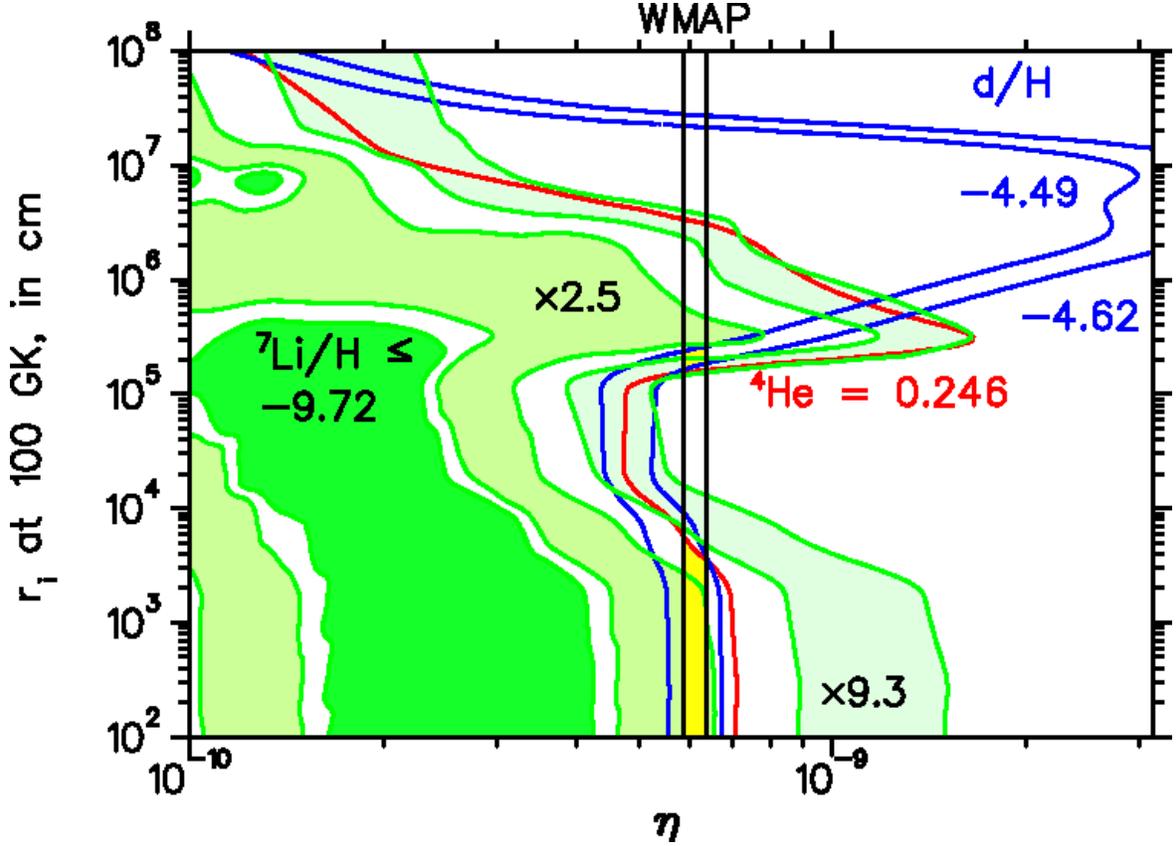}   
   \caption{\label{fig:li7Ry} ( Color Online ) Contours for the $\pm 2\sigma$ 
            abundance constraints on $ ^{7}$Li/H by Ryan et al. 
            \cite{rbofn00} are shown along with the constraints on deuterium 
            \cite{ktsol03}, CMB constraint from {\it WMAP} \cite{bennett03}, 
            and $+2\sigma$ maximum constraint on $ ^{4}$He \cite{it04} for the 
            case of $\tau_{\neu} = 878.5.\pm 0.7_{\stat} \pm 0.3_{\sys}$ 
            \cite{serebrov05}.  The region of concordance between {\it WMAP}, 
            $ ^{4}$He and deuterium is shown in yellow.  The contour lines for
            $ ^{7}$Li allow for depletion factors from 2.5 to 9.3 to bring 
            all three observational constraints and CMB constraints in 
            concordance with each other.}
\end{figure}

\begin{figure}
   \includegraphics[scale=0.80,angle=90]{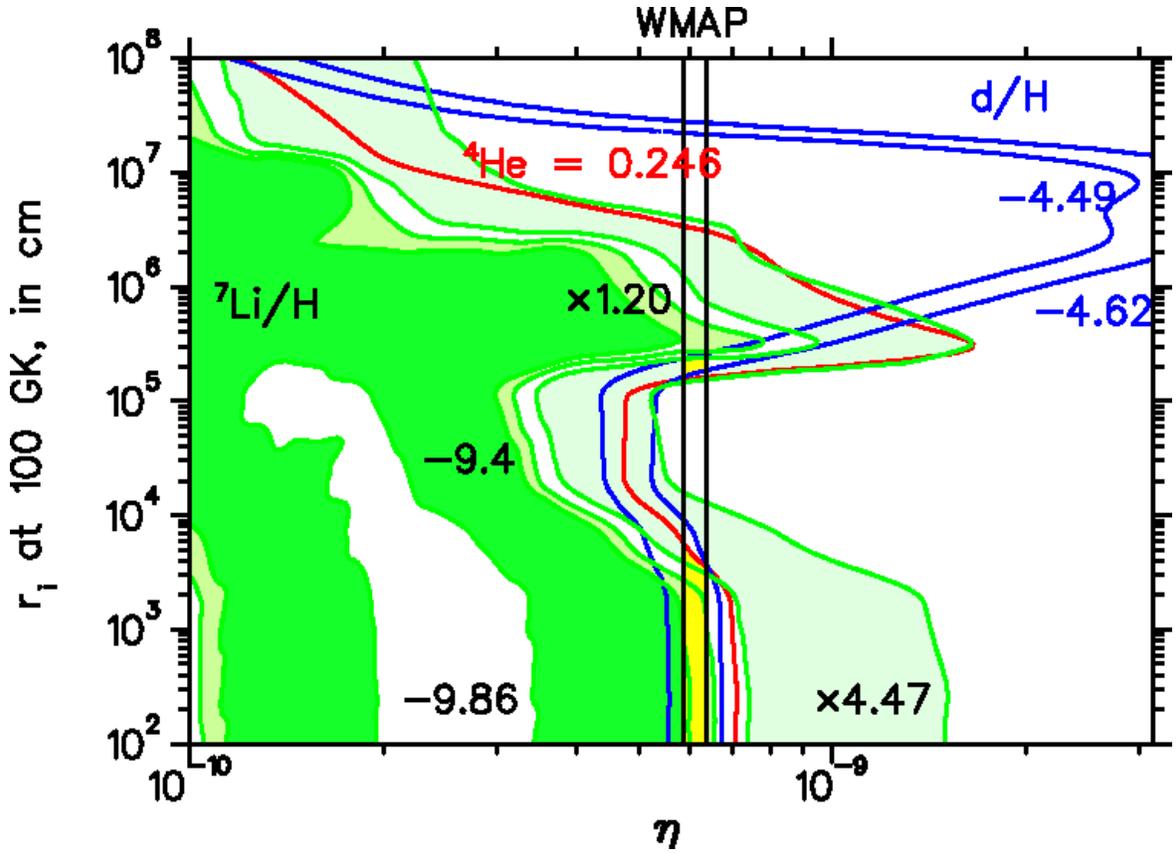}   
   \caption{\label{fig:li7MR} ( Color Online ) Same as in 
            Figure~\ref{fig:li7Ry} but with $\pm 2\sigma$ constraints on 
            $ ^{7}$Li/H by Melendez \& Ramirez \cite{mr04}. A thin 
            region of concordance between the $+2\sigma$ limit of the $ ^{7}$Li
            constraints and the other constraints exists for $r_{i} \le $ 1000 
            cm.  A depletion factor of 1.2 improves concordance.  A depletion 
            factor of 4.47 would bring these $ ^{7}$Li measurements in 
            concordance with the other constraints for 
            $r_{i} = (1.3-2.6) \times 10^{5}$ cm.}
\end{figure}

The primordial abundance of $ ^{7}$Li is in dispute because of disagreement 
over the calculation of the effective temperatures of metal poor halo stars 
needed to determine the abundances.  Ryan et al. \cite{rbofn00} and Melendez
\& Ramirez \cite{mr04} calculate different values of the effective temperature 
and different ranges of the $ ^{7}$Li abundance.   Figure~\ref{fig:li7Ry} shows
the Ryan et al. $ ^{7}$Li constraints $ ^{7}$Li/H 
$ = 1.23_{-0.32}^{+0.68} \times 10^{-10}$ along with the deuterium, {\it WMAP} 
and $ ^{4}$He constraints when using the Serebrov et al. lifetime.  
Figure~\ref{fig:li7MR} does the same but instead shows the Melendez \& Ramirez 
constraints $ ^{7}$Li/H = $ 2.34_{-0.96}^{+1.64} \times 10^{-10}$.  Regions of 
concordance between the $ ^{4}$He, deuterium, and {\it WMAP} constraints are 
highlighted in yellow (color online).

No region of concordance exists between the $ ^{7}$Li constraints of Ryan et 
al. \cite{rbofn00} and the constraints from deuterium and $ ^{4}$He for either
the SBBN model or the IBBN model.  A depletion factor due to stellar processes
is needed for concordance.  A factor of 2.5 would bring the $ ^{7}$Li 
constraints into concordance with the {\it WMAP} constraints 
$\eta = 6.13 \pm 0.25 \times 10^{-10}$ for the SBBN case and for the IBBN case
with distance scale $r_{i} \le $ 2500 cm.  But for $r_{i} \approx $ 25000 cm 
nucleosynthesis in high proton density regions leads to considerably increased
production of $ ^{7}$Be \cite{lara05}.  The $ ^{7}$Li contour lines shift to 
lower $\eta$ than the contour lines for deuterium and $ ^{4}$He.  Larger 
depletion factors can then be permitted in the IBBN model.  Depletion factors 
up to 5.9 are permitted using the world average neutron lifetime \cite{lara05}.
The main effect of using the Serebrov et al. lifetime is to shift the $ ^{4}$He
contour lines to higher $\eta$.  This shift increases the maximum allowed range
for the depletion factor, to 9.3 as shown in Figure~\ref{fig:li7Ry}.  The 
constraints by Ryan et al. with this depletion factor will be in concordance 
with the narrow region $r_{i} = (1.3-2.6) \times 10^{5}$ cm permitted by the
deuterium, $ ^{4}$He, and {\it WMAP} constraints.

Figure~\ref{fig:li7MR} shows a thin region of concordance 
$\eta = (5.88-6.0) \times 10^{-10}$ between the {\it WMAP} constraints and the 
2$\sigma$ limit of the $ ^{7}$Li constraints by Melendez \& Ramirez 
\cite{mr04}.  A depletion factor of 1.2 would bring these $ ^{7}$Li constraints
into concordance with the whole range $\eta = 6.13 \pm 0.25 \times 10^{-10}$ of
the {\it WMAP} constraints for SBBN and for IBBN with $r_{i} \le $ 3000 cm.  
Using the world average neutron lifetime a maximum depletion factor of 2.8 of 
the Melendez \& Ramirez constraints is required by this IBBN model 
\cite{lara05}.  Using the Serebrov et al. neutron lifetime, a maximum depletion
factor of 4.47 brings the Melendez \& Ramirez constraints with the region 
$r_{i} = (1.3-2.6) \times 10^{5}$ cm permitted by the deuterium, $ ^{4}$He, and
{\it WMAP} constraints.

The results in this article are for a cylindrically symmetric IBBN model with 
an initial low density core and high density outer region.  Changing the 
symmetry of the model or changing the values of parameters $R_{\rho}$ and 
$f_{v}$ have the effect of lessening or amplifying the turns in the contour 
lines, but the characteristics of the contour lines remain the same 
qualitively.  Contour lines shift to lower $\eta$ and then to higher $\eta$ and
then back to lower $\eta$ as the size of the model increases.  The particular 
model of this article is thus representative of IBBN models in general.

We note that new measurements of the $ ^{7}$Li primordial abundance derived 
from the ratio $(  ^{7}\li/^{6}\li )$ measured in the interstellar medium 
\cite{ksak03,kawanomoto05} support a larger range of permitted depletion 
factors, consistent with those required here, leaving an enhanced $ ^{9}$Be or
$ ^{11}$B abundance as a possible observable signature for the IBBN
\cite{bk89,mf89,kb90,kmf90,tsof93,vroc98}.

\section{Conclusions}

We have reanalyzed IBBN in the context of the latest constraints on primordial
abundances, input physics, and cosmological parameters.  We have shown that 
some possible IBBN solutions are possible.  For example, a cylindrical 
fluctuation with a comoving radius of $\sim 10^5$ cm fixed at a temperature of
100 GK is still consistent with the D/H, $Y_{p}$ and WMAP constraints.  
Depletion factors of up to 9.3 or 4.47, depending on which measurement is used,
can bring the $ ^{7}$Li results into concordance as well.  Also any fluctuation
up to a radius of $\sim 10^4$ cm is viable.  Hence, although the standard 
homogeneous BBN limit is still a consistent and the simplest solution, IBBN 
remains a viable possibility should it ever be established that such 
fluctuations arise from earlier cosmic phase transitions.

\section{Acknowledgements}

This work has been supported in part by Grants-in-Aid for Scientific
Research (14540271, 17540275) and for Specially Promoted Reseach (13002001)
of the Ministry of Education, Culture, Sports, Science and Technology of
Japan, and the Mitsubishi Foundation.  This work was also partially funded by 
National Science Foundation grants PHY 9800725, PHY 0102204, and PHY 035482.  
Work at UND has been supported under DoE Nuclear Theory grant \# 
DE-FG02-95-ER40934.  The authors thank the National Astronomical Observatory of
Japan and their respective academic institutions for the opportunity to carry 
out this research.  

\bibliography{PRD02}

\begin{thebibliography}{80}
\expandafter\ifx\csname natexlab\endcsname\relax\def\natexlab#1{#1}\fi
\expandafter\ifx\csname bibnamefont\endcsname\relax
  \def\bibnamefont#1{#1}\fi
\expandafter\ifx\csname bibfnamefont\endcsname\relax
  \def\bibfnamefont#1{#1}\fi
\expandafter\ifx\csname citenamefont\endcsname\relax
  \def\citenamefont#1{#1}\fi
\expandafter\ifx\csname url\endcsname\relax
  \def\url#1{\texttt{#1}}\fi
\expandafter\ifx\csname urlprefix\endcsname\relax\def\urlprefix{URL }\fi
\providecommand{\bibinfo}[2]{#2}
\providecommand{\eprint}[2][]{\url{#2}}

\bibitem[{\citenamefont{Wagoner}(1973)}]{wagoner73}
\bibinfo{author}{\bibfnamefont{R.}~\bibnamefont{Wagoner}},
  \bibinfo{journal}{\apj} \textbf{\bibinfo{volume}{179}}, \bibinfo{pages}{343}
  (\bibinfo{year}{1973}).

\bibitem[{\citenamefont{Yang et~al.}(1984)\citenamefont{Yang, Turner, Steigman,
  Schramm, and Olive}}]{ytsso84}
\bibinfo{author}{\bibfnamefont{J.}~\bibnamefont{Yang}},
  \bibinfo{author}{\bibfnamefont{M.}~\bibnamefont{Turner}},
  \bibinfo{author}{\bibfnamefont{G.}~\bibnamefont{Steigman}},
  \bibinfo{author}{\bibfnamefont{D.}~\bibnamefont{Schramm}}, \bibnamefont{and}
  \bibinfo{author}{\bibfnamefont{K.}~\bibnamefont{Olive}},
  \bibinfo{journal}{\apj} \textbf{\bibinfo{volume}{281}}, \bibinfo{pages}{493}
  (\bibinfo{year}{1984}).

\bibitem[{\citenamefont{Smith et~al.}(1993)\citenamefont{Smith, Kawano, and
  Malaney}}]{skm93}
\bibinfo{author}{\bibfnamefont{M.}~\bibnamefont{Smith}},
  \bibinfo{author}{\bibfnamefont{L.}~\bibnamefont{Kawano}}, \bibnamefont{and}
  \bibinfo{author}{\bibfnamefont{R.}~\bibnamefont{Malaney}},
  \bibinfo{journal}{Astrophys.~J.~Suppl.} \textbf{\bibinfo{volume}{85}},
  \bibinfo{pages}{219} (\bibinfo{year}{1993}).

\bibitem[{\citenamefont{Melchiorri}(2003)}]{melchiorri03}
\bibinfo{author}{\bibfnamefont{A.}~\bibnamefont{Melchiorri}}, in
  \emph{\bibinfo{booktitle}{Dark Matter in Astro- and Particle Physics DARK
  2002}}, edited by
  \bibinfo{editor}{\bibfnamefont{H.}~\bibnamefont{Klapdor-Kleingrothaus}}
  \bibnamefont{and} \bibinfo{editor}{\bibfnamefont{R.}~\bibnamefont{Viollier}}
  (\bibinfo{publisher}{Springer}, \bibinfo{year}{2003}), p.
  \bibinfo{pages}{101}, \bibinfo{note}{astro-ph/0204262}.

\bibitem[{\citenamefont{Bennett and et~al.}(2003)}]{bennett03}
\bibinfo{author}{\bibfnamefont{C.}~\bibnamefont{Bennett}} \bibnamefont{and}
  \bibinfo{author}{\bibnamefont{et~al.}},
  \bibinfo{journal}{Astrophys.~J.~Suppl.} \textbf{\bibinfo{volume}{148}},
  \bibinfo{pages}{1} (\bibinfo{year}{2003}).

\bibitem[{\citenamefont{Lara}(2005)}]{lara05}
\bibinfo{author}{\bibfnamefont{J.}~\bibnamefont{Lara}}, \bibinfo{journal}{\prd}
  \textbf{\bibinfo{volume}{72}}, \bibinfo{pages}{023509}
  (\bibinfo{year}{2005}).

\bibitem[{\citenamefont{Eidelman and et~al [~Particle Data
  Group~]}(2004)}]{eidelman04}
\bibinfo{author}{\bibfnamefont{S.}~\bibnamefont{Eidelman}} \bibnamefont{and}
  \bibinfo{author}{\bibnamefont{et~al [~Particle Data Group~]}},
  \bibinfo{journal}{Phys.~Lett.~B} \textbf{\bibinfo{volume}{592}},
  \bibinfo{pages}{1} (\bibinfo{year}{2004}).

\bibitem[{\citenamefont{Serebrov and et~al}(2005)}]{serebrov05}
\bibinfo{author}{\bibfnamefont{A.}~\bibnamefont{Serebrov}} \bibnamefont{and}
  \bibinfo{author}{\bibnamefont{et~al}}, \bibinfo{journal}{Phys.~Lett.~B}
  \textbf{\bibinfo{volume}{605}}, \bibinfo{pages}{72} (\bibinfo{year}{2005}).

\bibitem[{\citenamefont{Abele et~al.}(2002)\citenamefont{Abele,
  Astruc-Hoffmann, Bae$\beta$ler, Dubbers, Gl$\uum$ck, M$\uum$ller, and
  Nesivzhevsky}}]{aabdgmnrz02}
\bibinfo{author}{\bibfnamefont{H.}~\bibnamefont{Abele}},
  \bibinfo{author}{\bibfnamefont{M.}~\bibnamefont{Astruc-Hoffmann}},
  \bibinfo{author}{\bibfnamefont{S.}~\bibnamefont{Bae$\beta$ler}},
  \bibinfo{author}{\bibfnamefont{D.}~\bibnamefont{Dubbers}},
  \bibinfo{author}{\bibfnamefont{F.}~\bibnamefont{Gl$\uum$ck}},
  \bibinfo{author}{\bibfnamefont{U.}~\bibnamefont{M$\uum$ller}},
  \bibnamefont{and}
  \bibinfo{author}{\bibfnamefont{V.}~\bibnamefont{Nesivzhevsky}},
  \bibinfo{journal}{\prl} \textbf{\bibinfo{volume}{88}},
  \bibinfo{pages}{211801} (\bibinfo{year}{2002}).

\bibitem[{\citenamefont{Abele et~al.}(2004)\citenamefont{Abele, Barberio,
  Dubbers, Gl$\uum$ck, Hardy, Marciano, Serebrov, and Severijns}}]{abdghmss04}
\bibinfo{author}{\bibfnamefont{H.}~\bibnamefont{Abele}},
  \bibinfo{author}{\bibfnamefont{E.}~\bibnamefont{Barberio}},
  \bibinfo{author}{\bibfnamefont{D.}~\bibnamefont{Dubbers}},
  \bibinfo{author}{\bibfnamefont{F.}~\bibnamefont{Gl$\uum$ck}},
  \bibinfo{author}{\bibfnamefont{J.}~\bibnamefont{Hardy}},
  \bibinfo{author}{\bibfnamefont{W.}~\bibnamefont{Marciano}},
  \bibinfo{author}{\bibfnamefont{A.}~\bibnamefont{Serebrov}}, \bibnamefont{and}
  \bibinfo{author}{\bibfnamefont{N.}~\bibnamefont{Severijns}},
  \bibinfo{journal}{Eur.~Phys.~J.~C} \textbf{\bibinfo{volume}{33}},
  \bibinfo{pages}{1} (\bibinfo{year}{2004}).

\bibitem[{\citenamefont{Mathews et~al.}(2005)\citenamefont{Mathews, Kajino, and
  Shima}}]{mks05}
\bibinfo{author}{\bibfnamefont{G.}~\bibnamefont{Mathews}},
  \bibinfo{author}{\bibfnamefont{T.}~\bibnamefont{Kajino}}, \bibnamefont{and}
  \bibinfo{author}{\bibfnamefont{T.}~\bibnamefont{Shima}},
  \bibinfo{journal}{\prd} \textbf{\bibinfo{volume}{71}},
  \bibinfo{pages}{021302} (\bibinfo{year}{2005}).

\bibitem[{\citenamefont{Izotov et~al.}(1999)\citenamefont{Izotov, Chafhee,
  Foltz, Green, Guseva, and Thuan}}]{icfggt99}
\bibinfo{author}{\bibfnamefont{Y.}~\bibnamefont{Izotov}},
  \bibinfo{author}{\bibfnamefont{F.}~\bibnamefont{Chafhee}},
  \bibinfo{author}{\bibfnamefont{C.}~\bibnamefont{Foltz}},
  \bibinfo{author}{\bibfnamefont{R.}~\bibnamefont{Green}},
  \bibinfo{author}{\bibfnamefont{N.}~\bibnamefont{Guseva}}, \bibnamefont{and}
  \bibinfo{author}{\bibfnamefont{T.}~\bibnamefont{Thuan}},
  \bibinfo{journal}{\apj} \textbf{\bibinfo{volume}{527}}, \bibinfo{pages}{757}
  (\bibinfo{year}{1999}).

\bibitem[{\citenamefont{Kawano}(1992)}]{kawano92}
\bibinfo{author}{\bibfnamefont{L.}~\bibnamefont{Kawano}}
  (\bibinfo{year}{1992}), \bibinfo{note}{fermilab-Pub-92/04-A}.

\bibitem[{\citenamefont{Kawano}()}]{website}
\bibinfo{author}{\bibfnamefont{L.}~\bibnamefont{Kawano}},
  \bibinfo{note}{available at
  http://www-thphys.physics.ox.ac.uk/user/SubirSarkar/bbn.html}.

\bibitem[{\citenamefont{Kurki-Suonio}(1988)}]{kurkisuonio88}
\bibinfo{author}{\bibfnamefont{H.}~\bibnamefont{Kurki-Suonio}},
  \bibinfo{journal}{\prd} \textbf{\bibinfo{volume}{37}}, \bibinfo{pages}{2104}
  (\bibinfo{year}{1988}).

\bibitem[{\citenamefont{Malaney and Mathews}(1993)}]{mm93}
\bibinfo{author}{\bibfnamefont{R.}~\bibnamefont{Malaney}} \bibnamefont{and}
  \bibinfo{author}{\bibfnamefont{G.}~\bibnamefont{Mathews}},
  \bibinfo{journal}{Phys.~Rep.} \textbf{\bibinfo{volume}{229}},
  \bibinfo{pages}{145} (\bibinfo{year}{1993}).

\bibitem[{\citenamefont{Sumiyoshi et~al.}(1990)\citenamefont{Sumiyoshi, Kajino,
  Alcock, and Mathews}}]{skam90}
\bibinfo{author}{\bibfnamefont{K.}~\bibnamefont{Sumiyoshi}},
  \bibinfo{author}{\bibfnamefont{T.}~\bibnamefont{Kajino}},
  \bibinfo{author}{\bibfnamefont{C.}~\bibnamefont{Alcock}}, \bibnamefont{and}
  \bibinfo{author}{\bibfnamefont{G.}~\bibnamefont{Mathews}},
  \bibinfo{journal}{\prd} \textbf{\bibinfo{volume}{42}}, \bibinfo{pages}{3963}
  (\bibinfo{year}{1990}).

\bibitem[{\citenamefont{Witten}(1984)}]{witten84}
\bibinfo{author}{\bibfnamefont{E.}~\bibnamefont{Witten}},
  \bibinfo{journal}{\prd} \textbf{\bibinfo{volume}{30}}, \bibinfo{pages}{272}
  (\bibinfo{year}{1984}).

\bibitem[{\citenamefont{Alcock et~al.}(1987)\citenamefont{Alcock, Fuller, and
  Mathews}}]{afm87}
\bibinfo{author}{\bibfnamefont{C.}~\bibnamefont{Alcock}},
  \bibinfo{author}{\bibfnamefont{G.}~\bibnamefont{Fuller}}, \bibnamefont{and}
  \bibinfo{author}{\bibfnamefont{G.}~\bibnamefont{Mathews}},
  \bibinfo{journal}{\apj} \textbf{\bibinfo{volume}{320}}, \bibinfo{pages}{439}
  (\bibinfo{year}{1987}).

\bibitem[{\citenamefont{Terasawa and Sato}(1989)}]{ts89}
\bibinfo{author}{\bibfnamefont{N.}~\bibnamefont{Terasawa}} \bibnamefont{and}
  \bibinfo{author}{\bibfnamefont{K.}~\bibnamefont{Sato}},
  \bibinfo{journal}{\prd} \textbf{\bibinfo{volume}{39}}, \bibinfo{pages}{2893}
  (\bibinfo{year}{1989}).

\bibitem[{\citenamefont{Kim et~al.}(2001)\citenamefont{Kim, Lee, and
  Lee}}]{kll01}
\bibinfo{author}{\bibfnamefont{H.-I.} \bibnamefont{Kim}},
  \bibinfo{author}{\bibfnamefont{B.-H.} \bibnamefont{Lee}}, \bibnamefont{and}
  \bibinfo{author}{\bibfnamefont{C.-H.} \bibnamefont{Lee}},
  \bibinfo{journal}{\prd} \textbf{\bibinfo{volume}{64}},
  \bibinfo{pages}{067301} (\bibinfo{year}{2001}).

\bibitem[{\citenamefont{Banerjee and Gavai}(1992)}]{bg92}
\bibinfo{author}{\bibfnamefont{B.}~\bibnamefont{Banerjee}} \bibnamefont{and}
  \bibinfo{author}{\bibfnamefont{R.}~\bibnamefont{Gavai}},
  \bibinfo{journal}{Phys.~Lett.~B} \textbf{\bibinfo{volume}{293}},
  \bibinfo{pages}{157} (\bibinfo{year}{1992}).

\bibitem[{\citenamefont{Coley and Trappenberg}(1994)}]{ct93}
\bibinfo{author}{\bibfnamefont{A.}~\bibnamefont{Coley}} \bibnamefont{and}
  \bibinfo{author}{\bibfnamefont{T.}~\bibnamefont{Trappenberg}},
  \bibinfo{journal}{\prd} \textbf{\bibinfo{volume}{50}}, \bibinfo{pages}{4881}
  (\bibinfo{year}{1994}).

\bibitem[{\citenamefont{Fuller et~al.}(1994)\citenamefont{Fuller, Jedamzik,
  Mathews, and Olinto}}]{fjmo94}
\bibinfo{author}{\bibfnamefont{G.}~\bibnamefont{Fuller}},
  \bibinfo{author}{\bibfnamefont{K.}~\bibnamefont{Jedamzik}},
  \bibinfo{author}{\bibfnamefont{G.}~\bibnamefont{Mathews}}, \bibnamefont{and}
  \bibinfo{author}{\bibfnamefont{A.}~\bibnamefont{Olinto}},
  \bibinfo{journal}{Phys.~Lett.~B} \textbf{\bibinfo{volume}{333}},
  \bibinfo{pages}{135} (\bibinfo{year}{1994}).

\bibitem[{\citenamefont{Heckler}(1995)}]{heckler95}
\bibinfo{author}{\bibfnamefont{A.}~\bibnamefont{Heckler}},
  \bibinfo{journal}{\prd} \textbf{\bibinfo{volume}{51}}, \bibinfo{pages}{405}
  (\bibinfo{year}{1995}).

\bibitem[{\citenamefont{Kainulainen et~al.}(1999)\citenamefont{Kainulainen,
  Kurki-Suonio, and Sihvola}}]{kks99}
\bibinfo{author}{\bibfnamefont{K.}~\bibnamefont{Kainulainen}},
  \bibinfo{author}{\bibfnamefont{H.}~\bibnamefont{Kurki-Suonio}},
  \bibnamefont{and} \bibinfo{author}{\bibfnamefont{E.}~\bibnamefont{Sihvola}},
  \bibinfo{journal}{\prd} \textbf{\bibinfo{volume}{59}},
  \bibinfo{pages}{083505} (\bibinfo{year}{1999}).

\bibitem[{\citenamefont{M\'{e}gevand and Astorga}(2005)}]{ma05}
\bibinfo{author}{\bibfnamefont{A.}~\bibnamefont{M\'{e}gevand}}
  \bibnamefont{and} \bibinfo{author}{\bibfnamefont{F.}~\bibnamefont{Astorga}},
  \bibinfo{journal}{\prd} \textbf{\bibinfo{volume}{71}},
  \bibinfo{pages}{023502} (\bibinfo{year}{2005}).

\bibitem[{\citenamefont{Brandenberger et~al.}(1995)\citenamefont{Brandenberger,
  Davis, and Rees}}]{bdr95}
\bibinfo{author}{\bibfnamefont{R.}~\bibnamefont{Brandenberger}},
  \bibinfo{author}{\bibfnamefont{A.-C.} \bibnamefont{Davis}}, \bibnamefont{and}
  \bibinfo{author}{\bibfnamefont{M.}~\bibnamefont{Rees}},
  \bibinfo{journal}{Phys.~Lett.~B} \textbf{\bibinfo{volume}{349}},
  \bibinfo{pages}{329} (\bibinfo{year}{1995}).

\bibitem[{\citenamefont{Orito et~al.}(1997)\citenamefont{Orito, Kajino, Boyd,
  and Mathews}}]{okbm97}
\bibinfo{author}{\bibfnamefont{M.}~\bibnamefont{Orito}},
  \bibinfo{author}{\bibfnamefont{T.}~\bibnamefont{Kajino}},
  \bibinfo{author}{\bibfnamefont{R.}~\bibnamefont{Boyd}}, \bibnamefont{and}
  \bibinfo{author}{\bibfnamefont{G.}~\bibnamefont{Mathews}},
  \bibinfo{journal}{\apj} \textbf{\bibinfo{volume}{488}}, \bibinfo{pages}{515}
  (\bibinfo{year}{1997}).

\bibitem[{\citenamefont{Layek et~al.}(2001)\citenamefont{Layek, Sanyal, and
  Srivastava}}]{lss01}
\bibinfo{author}{\bibfnamefont{B.}~\bibnamefont{Layek}},
  \bibinfo{author}{\bibfnamefont{S.}~\bibnamefont{Sanyal}}, \bibnamefont{and}
  \bibinfo{author}{\bibfnamefont{A.}~\bibnamefont{Srivastava}},
  \bibinfo{journal}{\prd} \textbf{\bibinfo{volume}{63}},
  \bibinfo{pages}{083512} (\bibinfo{year}{2001}).

\bibitem[{\citenamefont{Layek et~al.}(2003)\citenamefont{Layek, Sanyal, and
  Srivastava}}]{lss03}
\bibinfo{author}{\bibfnamefont{B.}~\bibnamefont{Layek}},
  \bibinfo{author}{\bibfnamefont{S.}~\bibnamefont{Sanyal}}, \bibnamefont{and}
  \bibinfo{author}{\bibfnamefont{A.}~\bibnamefont{Srivastava}},
  \bibinfo{journal}{\prd} \textbf{\bibinfo{volume}{67}},
  \bibinfo{pages}{083508} (\bibinfo{year}{2003}).

\bibitem[{\citenamefont{Applegate et~al.}(1987)\citenamefont{Applegate, Hogan,
  and Scherrer}}]{ahs87}
\bibinfo{author}{\bibfnamefont{J.}~\bibnamefont{Applegate}},
  \bibinfo{author}{\bibfnamefont{C.}~\bibnamefont{Hogan}}, \bibnamefont{and}
  \bibinfo{author}{\bibfnamefont{R.}~\bibnamefont{Scherrer}},
  \bibinfo{journal}{\prd} \textbf{\bibinfo{volume}{35}}, \bibinfo{pages}{1151}
  (\bibinfo{year}{1987}).

\bibitem[{\citenamefont{Malaney and Fowler}(1988)}]{mf88}
\bibinfo{author}{\bibfnamefont{R.}~\bibnamefont{Malaney}} \bibnamefont{and}
  \bibinfo{author}{\bibfnamefont{W.}~\bibnamefont{Fowler}},
  \bibinfo{journal}{\apj} \textbf{\bibinfo{volume}{333}}, \bibinfo{pages}{14}
  (\bibinfo{year}{1988}).

\bibitem[{\citenamefont{Kurki-Suonio et~al.}(1988)\citenamefont{Kurki-Suonio,
  Matzner, Centrella, Rothman, and Wilson}}]{kmcrw88}
\bibinfo{author}{\bibfnamefont{H.}~\bibnamefont{Kurki-Suonio}},
  \bibinfo{author}{\bibfnamefont{R.}~\bibnamefont{Matzner}},
  \bibinfo{author}{\bibfnamefont{J.}~\bibnamefont{Centrella}},
  \bibinfo{author}{\bibfnamefont{T.}~\bibnamefont{Rothman}}, \bibnamefont{and}
  \bibinfo{author}{\bibfnamefont{J.}~\bibnamefont{Wilson}},
  \bibinfo{journal}{\prd} \textbf{\bibinfo{volume}{38}}, \bibinfo{pages}{1091}
  (\bibinfo{year}{1988}).

\bibitem[{\citenamefont{Mathews et~al.}(1990)\citenamefont{Mathews, Meyer,
  Alcock, and Fuller}}]{mmaf90}
\bibinfo{author}{\bibfnamefont{G.}~\bibnamefont{Mathews}},
  \bibinfo{author}{\bibfnamefont{B.}~\bibnamefont{Meyer}},
  \bibinfo{author}{\bibfnamefont{C.}~\bibnamefont{Alcock}}, \bibnamefont{and}
  \bibinfo{author}{\bibfnamefont{G.}~\bibnamefont{Fuller}},
  \bibinfo{journal}{\apj} \textbf{\bibinfo{volume}{358}}, \bibinfo{pages}{36}
  (\bibinfo{year}{1990}).

\bibitem[{\citenamefont{Zeldovich}(1975)}]{zeldovich75}
\bibinfo{author}{\bibfnamefont{I.}~\bibnamefont{Zeldovich}},
  \bibinfo{journal}{Sov.~Astr.~Lett.} \textbf{\bibinfo{volume}{1}},
  \bibinfo{pages}{5} (\bibinfo{year}{1975}).

\bibitem[{\citenamefont{Epstein and Petrosian}(1975)}]{ep75}
\bibinfo{author}{\bibfnamefont{R.}~\bibnamefont{Epstein}} \bibnamefont{and}
  \bibinfo{author}{\bibfnamefont{V.}~\bibnamefont{Petrosian}},
  \bibinfo{journal}{\apj} \textbf{\bibinfo{volume}{197}}, \bibinfo{pages}{281}
  (\bibinfo{year}{1975}).

\bibitem[{\citenamefont{Hogan}(1978)}]{hogan78}
\bibinfo{author}{\bibfnamefont{C.}~\bibnamefont{Hogan}},
  \bibinfo{journal}{Mon.~Not.~R.~Astron.~Soc.} \textbf{\bibinfo{volume}{185}},
  \bibinfo{pages}{889} (\bibinfo{year}{1978}).

\bibitem[{\citenamefont{Barrow and Morgan}(1983)}]{bm83}
\bibinfo{author}{\bibfnamefont{J.}~\bibnamefont{Barrow}} \bibnamefont{and}
  \bibinfo{author}{\bibfnamefont{J.}~\bibnamefont{Morgan}},
  \bibinfo{journal}{Mon.~Not.~R.~Astron.~Soc.} \textbf{\bibinfo{volume}{203}},
  \bibinfo{pages}{393} (\bibinfo{year}{1983}).

\bibitem[{\citenamefont{Sale and Mathews}(1986)}]{sm86}
\bibinfo{author}{\bibfnamefont{K.}~\bibnamefont{Sale}} \bibnamefont{and}
  \bibinfo{author}{\bibfnamefont{G.}~\bibnamefont{Mathews}},
  \bibinfo{journal}{\apj} \textbf{\bibinfo{volume}{309}}, \bibinfo{pages}{L1}
  (\bibinfo{year}{1986}).

\bibitem[{\citenamefont{Kurki-Suonio and Matzner}(1989)}]{km89}
\bibinfo{author}{\bibfnamefont{H.}~\bibnamefont{Kurki-Suonio}}
  \bibnamefont{and} \bibinfo{author}{\bibfnamefont{R.}~\bibnamefont{Matzner}},
  \bibinfo{journal}{\prd} \textbf{\bibinfo{volume}{39}}, \bibinfo{pages}{1046}
  (\bibinfo{year}{1989}).

\bibitem[{\citenamefont{Kurki-Suonio and Matzner}(1990)}]{km90}
\bibinfo{author}{\bibfnamefont{H.}~\bibnamefont{Kurki-Suonio}}
  \bibnamefont{and} \bibinfo{author}{\bibfnamefont{R.}~\bibnamefont{Matzner}},
  \bibinfo{journal}{\prd} \textbf{\bibinfo{volume}{42}}, \bibinfo{pages}{1047}
  (\bibinfo{year}{1990}).

\bibitem[{\citenamefont{Kurki-Suonio et~al.}(1990)\citenamefont{Kurki-Suonio,
  Matzner, Olive, and Schramm}}]{kmos90}
\bibinfo{author}{\bibfnamefont{H.}~\bibnamefont{Kurki-Suonio}},
  \bibinfo{author}{\bibfnamefont{R.}~\bibnamefont{Matzner}},
  \bibinfo{author}{\bibfnamefont{K.}~\bibnamefont{Olive}}, \bibnamefont{and}
  \bibinfo{author}{\bibfnamefont{D.}~\bibnamefont{Schramm}},
  \bibinfo{journal}{\apj} \textbf{\bibinfo{volume}{353}}, \bibinfo{pages}{406}
  (\bibinfo{year}{1990}).

\bibitem[{\citenamefont{Kajino et~al.}(1990)\citenamefont{Kajino, Mathews, and
  Fuller}}]{kmf90}
\bibinfo{author}{\bibfnamefont{T.}~\bibnamefont{Kajino}},
  \bibinfo{author}{\bibfnamefont{G.}~\bibnamefont{Mathews}}, \bibnamefont{and}
  \bibinfo{author}{\bibfnamefont{G.}~\bibnamefont{Fuller}},
  \bibinfo{journal}{\apj} \textbf{\bibinfo{volume}{364}}, \bibinfo{pages}{7}
  (\bibinfo{year}{1990}).

\bibitem[{\citenamefont{Meyers et~al.}(1991)\citenamefont{Meyers, Alcock,
  Mathews, and Fuller}}]{mamf91}
\bibinfo{author}{\bibfnamefont{B.}~\bibnamefont{Meyers}},
  \bibinfo{author}{\bibfnamefont{C.}~\bibnamefont{Alcock}},
  \bibinfo{author}{\bibfnamefont{G.}~\bibnamefont{Mathews}}, \bibnamefont{and}
  \bibinfo{author}{\bibfnamefont{G.}~\bibnamefont{Fuller}},
  \bibinfo{journal}{\prd} \textbf{\bibinfo{volume}{43}}, \bibinfo{pages}{1079}
  (\bibinfo{year}{1991}).

\bibitem[{\citenamefont{Thomas et~al.}(1994)\citenamefont{Thomas, D.N.,
  Schramm, Olive, Mathews, Meyer, and Fields}}]{tsommf94}
\bibinfo{author}{\bibfnamefont{D.}~\bibnamefont{Thomas}},
  \bibinfo{author}{\bibnamefont{D.N.}},
  \bibinfo{author}{\bibnamefont{Schramm}},
  \bibinfo{author}{\bibfnamefont{K.}~\bibnamefont{Olive}},
  \bibinfo{author}{\bibfnamefont{G.}~\bibnamefont{Mathews}},
  \bibinfo{author}{\bibfnamefont{B.}~\bibnamefont{Meyer}}, \bibnamefont{and}
  \bibinfo{author}{\bibfnamefont{B.}~\bibnamefont{Fields}},
  \bibinfo{journal}{\apj} \textbf{\bibinfo{volume}{430}}, \bibinfo{pages}{291}
  (\bibinfo{year}{1994}).

\bibitem[{\citenamefont{Jedamzik
  et~al.}(1994{\natexlab{a}})\citenamefont{Jedamzik, Fuller, Mathews, and
  Kajino}}]{jfmk94}
\bibinfo{author}{\bibfnamefont{K.}~\bibnamefont{Jedamzik}},
  \bibinfo{author}{\bibfnamefont{G.}~\bibnamefont{Fuller}},
  \bibinfo{author}{\bibfnamefont{G.}~\bibnamefont{Mathews}}, \bibnamefont{and}
  \bibinfo{author}{\bibfnamefont{T.}~\bibnamefont{Kajino}},
  \bibinfo{journal}{\apj} \textbf{\bibinfo{volume}{422}}, \bibinfo{pages}{423}
  (\bibinfo{year}{1994}{\natexlab{a}}).

\bibitem[{\citenamefont{Jedamzik
  et~al.}(1994{\natexlab{b}})\citenamefont{Jedamzik, Fuller, and
  Mathews}}]{jfm94}
\bibinfo{author}{\bibfnamefont{K.}~\bibnamefont{Jedamzik}},
  \bibinfo{author}{\bibfnamefont{G.}~\bibnamefont{Fuller}}, \bibnamefont{and}
  \bibinfo{author}{\bibfnamefont{G.}~\bibnamefont{Mathews}},
  \bibinfo{journal}{\apj} \textbf{\bibinfo{volume}{423}}, \bibinfo{pages}{50}
  (\bibinfo{year}{1994}{\natexlab{b}}).

\bibitem[{\citenamefont{Jedamzik et~al.}(1995)\citenamefont{Jedamzik, Mathews,
  and Fuller}}]{jmf95}
\bibinfo{author}{\bibfnamefont{K.}~\bibnamefont{Jedamzik}},
  \bibinfo{author}{\bibfnamefont{G.}~\bibnamefont{Mathews}}, \bibnamefont{and}
  \bibinfo{author}{\bibfnamefont{G.}~\bibnamefont{Fuller}},
  \bibinfo{journal}{\apj} \textbf{\bibinfo{volume}{441}}, \bibinfo{pages}{465}
  (\bibinfo{year}{1995}).

\bibitem[{\citenamefont{Leonard and Scherrer}(1996)}]{ls96}
\bibinfo{author}{\bibfnamefont{R.}~\bibnamefont{Leonard}} \bibnamefont{and}
  \bibinfo{author}{\bibfnamefont{R.}~\bibnamefont{Scherrer}},
  \bibinfo{journal}{\apj} \textbf{\bibinfo{volume}{463}}, \bibinfo{pages}{420}
  (\bibinfo{year}{1996}).

\bibitem[{\citenamefont{Mathews et~al.}(1996)\citenamefont{Mathews, Kajino, and
  Orito}}]{mko96}
\bibinfo{author}{\bibfnamefont{G.}~\bibnamefont{Mathews}},
  \bibinfo{author}{\bibfnamefont{T.}~\bibnamefont{Kajino}}, \bibnamefont{and}
  \bibinfo{author}{\bibfnamefont{M.}~\bibnamefont{Orito}},
  \bibinfo{journal}{\apj} \textbf{\bibinfo{volume}{456}}, \bibinfo{pages}{98}
  (\bibinfo{year}{1996}).

\bibitem[{\citenamefont{Kurki-Suonio et~al.}(1997)\citenamefont{Kurki-Suonio,
  Jedamzik, and Mathews}}]{kjm97}
\bibinfo{author}{\bibfnamefont{H.}~\bibnamefont{Kurki-Suonio}},
  \bibinfo{author}{\bibfnamefont{K.}~\bibnamefont{Jedamzik}}, \bibnamefont{and}
  \bibinfo{author}{\bibfnamefont{G.}~\bibnamefont{Mathews}},
  \bibinfo{journal}{\apj} \textbf{\bibinfo{volume}{479}}, \bibinfo{pages}{31}
  (\bibinfo{year}{1997}).

\bibitem[{\citenamefont{Suh and Mathews}(1998)}]{sm98}
\bibinfo{author}{\bibfnamefont{I.-S.} \bibnamefont{Suh}} \bibnamefont{and}
  \bibinfo{author}{\bibfnamefont{G.}~\bibnamefont{Mathews}},
  \bibinfo{journal}{\prd} \textbf{\bibinfo{volume}{58}},
  \bibinfo{pages}{123002} (\bibinfo{year}{1998}).

\bibitem[{\citenamefont{Ignatius and Schwarz}(2001)}]{is01}
\bibinfo{author}{\bibfnamefont{J.}~\bibnamefont{Ignatius}} \bibnamefont{and}
  \bibinfo{author}{\bibfnamefont{D.}~\bibnamefont{Schwarz}},
  \bibinfo{journal}{\prl} \textbf{\bibinfo{volume}{86}}, \bibinfo{pages}{2216}
  (\bibinfo{year}{2001}).

\bibitem[{\citenamefont{Jedamzik and Rehm}(2001)}]{jr01}
\bibinfo{author}{\bibfnamefont{K.}~\bibnamefont{Jedamzik}} \bibnamefont{and}
  \bibinfo{author}{\bibfnamefont{J.}~\bibnamefont{Rehm}},
  \bibinfo{journal}{\prd} \textbf{\bibinfo{volume}{64}},
  \bibinfo{pages}{023510} (\bibinfo{year}{2001}).

\bibitem[{\citenamefont{Kurki-Suonio and Sihvola}(2001)}]{ks01}
\bibinfo{author}{\bibfnamefont{H.}~\bibnamefont{Kurki-Suonio}}
  \bibnamefont{and} \bibinfo{author}{\bibfnamefont{E.}~\bibnamefont{Sihvola}},
  \bibinfo{journal}{\prd} \textbf{\bibinfo{volume}{63}},
  \bibinfo{pages}{083508} (\bibinfo{year}{2001}).

\bibitem[{\citenamefont{Kajino}(2002)}]{kajino02}
\bibinfo{author}{\bibfnamefont{T.}~\bibnamefont{Kajino}},
  \bibinfo{journal}{J.~Nucl.~Sci.~\& Tech.}
  \textbf{\bibinfo{volume}{Supp.~II}}, \bibinfo{pages}{530}
  (\bibinfo{year}{2002}).

\bibitem[{\citenamefont{Keihanen}(2002)}]{keihanen02}
\bibinfo{author}{\bibfnamefont{E.}~\bibnamefont{Keihanen}},
  \bibinfo{journal}{\prd} \textbf{\bibinfo{volume}{66}},
  \bibinfo{pages}{043512} (\bibinfo{year}{2002}).

\bibitem[{\citenamefont{Matsuura et~al.}(2004)\citenamefont{Matsuura, Dolgov,
  S.Nagataki, and Sato}}]{mdns04}
\bibinfo{author}{\bibfnamefont{S.}~\bibnamefont{Matsuura}},
  \bibinfo{author}{\bibfnamefont{A.}~\bibnamefont{Dolgov}},
  \bibinfo{author}{\bibnamefont{S.Nagataki}}, \bibnamefont{and}
  \bibinfo{author}{\bibfnamefont{K.}~\bibnamefont{Sato}},
  \bibinfo{journal}{Prog.~Theor.~Phys.} \textbf{\bibinfo{volume}{112}},
  \bibinfo{pages}{971} (\bibinfo{year}{2004}).

\bibitem[{\citenamefont{Matsuura et~al.}(2005)\citenamefont{Matsuura, Fujimoto,
  Nichimura, Hashimoto, and Sato}}]{mfnhs05}
\bibinfo{author}{\bibfnamefont{S.}~\bibnamefont{Matsuura}},
  \bibinfo{author}{\bibfnamefont{S.}~\bibnamefont{Fujimoto}},
  \bibinfo{author}{\bibfnamefont{S.}~\bibnamefont{Nichimura}},
  \bibinfo{author}{\bibfnamefont{M.}~\bibnamefont{Hashimoto}},
  \bibnamefont{and} \bibinfo{author}{\bibfnamefont{K.}~\bibnamefont{Sato}},
  \bibinfo{journal}{\prd} \textbf{\bibinfo{volume}{72}},
  \bibinfo{pages}{123505} (\bibinfo{year}{2005}).

\bibitem[{\citenamefont{Banerjee and Chitre}(1991)}]{bc91}
\bibinfo{author}{\bibfnamefont{B.}~\bibnamefont{Banerjee}} \bibnamefont{and}
  \bibinfo{author}{\bibfnamefont{S.}~\bibnamefont{Chitre}},
  \bibinfo{journal}{Phys.~Lett.~B} \textbf{\bibinfo{volume}{258}},
  \bibinfo{pages}{247} (\bibinfo{year}{1991}).

\bibitem[{\citenamefont{Kurki-Suonio et~al.}(1992)\citenamefont{Kurki-Suonio,
  Aufderheide, Graziani, Mathews, Banerjee, Chitre, and Schramm}}]{kagmbcs92}
\bibinfo{author}{\bibfnamefont{H.}~\bibnamefont{Kurki-Suonio}},
  \bibinfo{author}{\bibfnamefont{M.}~\bibnamefont{Aufderheide}},
  \bibinfo{author}{\bibfnamefont{F.}~\bibnamefont{Graziani}},
  \bibinfo{author}{\bibfnamefont{G.}~\bibnamefont{Mathews}},
  \bibinfo{author}{\bibfnamefont{B.}~\bibnamefont{Banerjee}},
  \bibinfo{author}{\bibfnamefont{S.}~\bibnamefont{Chitre}}, \bibnamefont{and}
  \bibinfo{author}{\bibfnamefont{D.}~\bibnamefont{Schramm}},
  \bibinfo{journal}{Phys.~Lett.~B} \textbf{\bibinfo{volume}{289}},
  \bibinfo{pages}{211} (\bibinfo{year}{1992}).

\bibitem[{\citenamefont{Jedamzik}()}]{jedamzik04}
\bibinfo{author}{\bibfnamefont{K.}~\bibnamefont{Jedamzik}},
  \bibinfo{note}{personal correspondence}.

\bibitem[{\citenamefont{Angulo and et~al.}(1999)}]{angulo99}
\bibinfo{author}{\bibfnamefont{C.}~\bibnamefont{Angulo}} \bibnamefont{and}
  \bibinfo{author}{\bibnamefont{et~al.}}, \bibinfo{journal}{Nucl.~Phys.~A}
  \textbf{\bibinfo{volume}{656}}, \bibinfo{pages}{3} (\bibinfo{year}{1999}).

\bibitem[{\citenamefont{Lara}(2004)}]{lara04}
\bibinfo{author}{\bibfnamefont{J.}~\bibnamefont{Lara}}, in
  \emph{\bibinfo{booktitle}{Frontier in Astroparticle Physics and Cosmology}},
  edited by \bibinfo{editor}{\bibfnamefont{K.}~\bibnamefont{Sato}}
  \bibnamefont{and} \bibinfo{editor}{\bibfnamefont{S.}~\bibnamefont{Nagataki}},
  \bibinfo{organization}{Research Center for the Early Universe}
  (\bibinfo{publisher}{Universal Academy Press}, \bibinfo{year}{2004}),
  p.~\bibinfo{pages}{87}, \bibinfo{note}{astro-ph/0402112}.

\bibitem[{\citenamefont{Weinberg}(1972)}]{weinberg72}
\bibinfo{author}{\bibfnamefont{S.}~\bibnamefont{Weinberg}},
  \emph{\bibinfo{title}{Graviation and Cosmology}} (\bibinfo{publisher}{John
  Wiley and Sons Inc.}, \bibinfo{year}{1972}), p. \bibinfo{pages}{547}.

\bibitem[{\citenamefont{Lara}(2001)}]{lara01}
\bibinfo{author}{\bibfnamefont{J.}~\bibnamefont{Lara}}, Ph.D. thesis,
  \bibinfo{school}{University of Texas at Austin} (\bibinfo{year}{2001}).

\bibitem[{\citenamefont{Dicus et~al.}(1982)\citenamefont{Dicus, Kolb, Gleeson,
  Sudarshan, Teplitz, and Turner}}]{dkgstt82}
\bibinfo{author}{\bibfnamefont{D.}~\bibnamefont{Dicus}},
  \bibinfo{author}{\bibfnamefont{E.}~\bibnamefont{Kolb}},
  \bibinfo{author}{\bibfnamefont{A.}~\bibnamefont{Gleeson}},
  \bibinfo{author}{\bibfnamefont{E.}~\bibnamefont{Sudarshan}},
  \bibinfo{author}{\bibfnamefont{V.}~\bibnamefont{Teplitz}}, \bibnamefont{and}
  \bibinfo{author}{\bibfnamefont{M.}~\bibnamefont{Turner}},
  \bibinfo{journal}{\prd} \textbf{\bibinfo{volume}{26}}, \bibinfo{pages}{2694}
  (\bibinfo{year}{1982}).

\bibitem[{\citenamefont{Kirkman et~al.}(2003)\citenamefont{Kirkman, Tytler,
  Suzuki, O'Meara, and Lubin}}]{ktsol03}
\bibinfo{author}{\bibfnamefont{D.}~\bibnamefont{Kirkman}},
  \bibinfo{author}{\bibfnamefont{D.}~\bibnamefont{Tytler}},
  \bibinfo{author}{\bibfnamefont{N.}~\bibnamefont{Suzuki}},
  \bibinfo{author}{\bibfnamefont{J.}~\bibnamefont{O'Meara}}, \bibnamefont{and}
  \bibinfo{author}{\bibfnamefont{D.}~\bibnamefont{Lubin}},
  \bibinfo{journal}{Astrophys.~J.~Suppl.} \textbf{\bibinfo{volume}{149}},
  \bibinfo{pages}{1} (\bibinfo{year}{2003}).

\bibitem[{\citenamefont{Izotov and Thuan}(2004)}]{it04}
\bibinfo{author}{\bibfnamefont{Y.}~\bibnamefont{Izotov}} \bibnamefont{and}
  \bibinfo{author}{\bibfnamefont{T.}~\bibnamefont{Thuan}},
  \bibinfo{journal}{\apj} \textbf{\bibinfo{volume}{602}}, \bibinfo{pages}{200}
  (\bibinfo{year}{2004}).

\bibitem[{\citenamefont{Olive and Skillman}(2004)}]{os04}
\bibinfo{author}{\bibfnamefont{K.}~\bibnamefont{Olive}} \bibnamefont{and}
  \bibinfo{author}{\bibfnamefont{E.}~\bibnamefont{Skillman}},
  \bibinfo{journal}{\apj} \textbf{\bibinfo{volume}{617}}, \bibinfo{pages}{29}
  (\bibinfo{year}{2004}).

\bibitem[{\citenamefont{Ryan et~al.}(2000)\citenamefont{Ryan, Beers, Olive,
  Fields, and Norris}}]{rbofn00}
\bibinfo{author}{\bibfnamefont{S.}~\bibnamefont{Ryan}},
  \bibinfo{author}{\bibfnamefont{T.}~\bibnamefont{Beers}},
  \bibinfo{author}{\bibfnamefont{K.}~\bibnamefont{Olive}},
  \bibinfo{author}{\bibfnamefont{B.}~\bibnamefont{Fields}}, \bibnamefont{and}
  \bibinfo{author}{\bibfnamefont{J.}~\bibnamefont{Norris}},
  \bibinfo{journal}{\apj} \textbf{\bibinfo{volume}{530}}, \bibinfo{pages}{L57}
  (\bibinfo{year}{2000}).

\bibitem[{\citenamefont{Melendez and Ramirez}(2004)}]{mr04}
\bibinfo{author}{\bibfnamefont{J.}~\bibnamefont{Melendez}} \bibnamefont{and}
  \bibinfo{author}{\bibfnamefont{I.}~\bibnamefont{Ramirez}},
  \bibinfo{journal}{\apj} \textbf{\bibinfo{volume}{615}}, \bibinfo{pages}{L33}
  (\bibinfo{year}{2004}).

\bibitem[{\citenamefont{Kawanomoto et~al.}(2003)\citenamefont{Kawanomoto,
  Suzuki, Ando, and Kajino}}]{ksak03}
\bibinfo{author}{\bibfnamefont{S.}~\bibnamefont{Kawanomoto}},
  \bibinfo{author}{\bibfnamefont{T.-K.} \bibnamefont{Suzuki}},
  \bibinfo{author}{\bibfnamefont{H.}~\bibnamefont{Ando}}, \bibnamefont{and}
  \bibinfo{author}{\bibfnamefont{T.}~\bibnamefont{Kajino}},
  \bibinfo{journal}{Nucl.~Phys.~A} \textbf{\bibinfo{volume}{718}},
  \bibinfo{pages}{659} (\bibinfo{year}{2003}).

\bibitem[{\citenamefont{Kawanomoto and et~al. [~SUBARU/HDS
  Collaboration~]}(2005)}]{kawanomoto05}
\bibinfo{author}{\bibfnamefont{S.}~\bibnamefont{Kawanomoto}} \bibnamefont{and}
  \bibinfo{author}{\bibnamefont{et~al. [~SUBARU/HDS Collaboration~]}}
  (\bibinfo{year}{2005}), \bibinfo{note}{submittal to Astrophys. J}.

\bibitem[{\citenamefont{Boyd and Kajino}(1989)}]{bk89}
\bibinfo{author}{\bibfnamefont{R.}~\bibnamefont{Boyd}} \bibnamefont{and}
  \bibinfo{author}{\bibfnamefont{T.}~\bibnamefont{Kajino}},
  \bibinfo{journal}{\apj} \textbf{\bibinfo{volume}{336}}, \bibinfo{pages}{L55}
  (\bibinfo{year}{1989}).

\bibitem[{\citenamefont{Malaney and Fowler}(1989)}]{mf89}
\bibinfo{author}{\bibfnamefont{R.}~\bibnamefont{Malaney}} \bibnamefont{and}
  \bibinfo{author}{\bibfnamefont{W.}~\bibnamefont{Fowler}},
  \bibinfo{journal}{\apj} \textbf{\bibinfo{volume}{345}}, \bibinfo{pages}{L5}
  (\bibinfo{year}{1989}).

\bibitem[{\citenamefont{Kajino and Boyd}(1990)}]{kb90}
\bibinfo{author}{\bibfnamefont{T.}~\bibnamefont{Kajino}} \bibnamefont{and}
  \bibinfo{author}{\bibfnamefont{R.}~\bibnamefont{Boyd}},
  \bibinfo{journal}{\apj} \textbf{\bibinfo{volume}{359}}, \bibinfo{pages}{267}
  (\bibinfo{year}{1990}).

\bibitem[{\citenamefont{Thomas et~al.}(1993)\citenamefont{Thomas, D.N.,
  Schramm, Olive, and Fields}}]{tsof93}
\bibinfo{author}{\bibfnamefont{D.}~\bibnamefont{Thomas}},
  \bibinfo{author}{\bibnamefont{D.N.}},
  \bibinfo{author}{\bibnamefont{Schramm}},
  \bibinfo{author}{\bibfnamefont{K.}~\bibnamefont{Olive}}, \bibnamefont{and}
  \bibinfo{author}{\bibfnamefont{B.}~\bibnamefont{Fields}},
  \bibinfo{journal}{\apj} \textbf{\bibinfo{volume}{406}}, \bibinfo{pages}{569}
  (\bibinfo{year}{1993}).

\bibitem[{\citenamefont{Vangioni-Flam et~al.}(1998)\citenamefont{Vangioni-Flam,
  Ramaty, Olive, and Cass\'{e}}}]{vroc98}
\bibinfo{author}{\bibfnamefont{E.}~\bibnamefont{Vangioni-Flam}},
  \bibinfo{author}{\bibfnamefont{R.}~\bibnamefont{Ramaty}},
  \bibinfo{author}{\bibfnamefont{K.}~\bibnamefont{Olive}}, \bibnamefont{and}
  \bibinfo{author}{\bibfnamefont{M.}~\bibnamefont{Cass\'{e}}},
  \bibinfo{journal}{Astron.~\& Astrophys.} \textbf{\bibinfo{volume}{337}},
  \bibinfo{pages}{714} (\bibinfo{year}{1998}).

\end{thebibliography}

\clearpage

\clearpage

\clearpage

\clearpage

\end{document}